\documentclass[10pt]{article}

\usepackage{amsmath}
\usepackage{amssymb}

\usepackage{graphicx}

\usepackage{cite}

\usepackage{color} 

\usepackage{footnote}
\usepackage{threeparttable}

\usepackage[labelfont=bf,labelsep=period,justification=raggedright]{caption}

\makeatletter
\renewcommand{\@biblabel}[1]{\quad#1.}
\makeatother

\date{}

\let\oldsqrt\sqrt
\def\sqrt{\mathpalette\DHLhksqrt}
\def\DHLhksqrt#1#2{%
    \setbox0=\hbox{$#1\oldsqrt{#2\,}$}\dimen0=\ht0
    \advance\dimen0-0.2\ht0
    \setbox2=\hbox{\vrule height\ht0 depth -\dimen0}%
{\box0\lower0.4pt\box2}}

\newcommand{\slfrac}[2]{\left.#1\middle/#2\right.}
\begin{document}

% Title must be 150 characters or less
\begin{flushleft}
{\Large
\textbf{Protein structure validation and refinement using amide proton chemical shifts derived from quantum mechanics}
}
% Insert Author names, affiliations and corresponding author email.
\\
Anders S. Christensen$^{1,\ast}$, 
Troels E. Linnet$^{2}$, 
Mikael Borg$^{3}$
Wouter Boomsma$^{2}$
Kresten Lindorff-Larsen$^{2}$
Thomas Hamelryck$^{3}$
Jan H. Jensen$^{1}$
\\
\bf{1} Department of Chemistry, University of Copenhagen, Copenhagen, Denmark
\\
\bf{2} Structural Biology and NMR Laboratory, Department of Biology, University of Copenhagen, Copenhagen, Denmark
\\
\bf{3} Structural Bioinformatics group, Section for Computational and RNA Biology, Department of Biology, University of Copenhagen, Copenhagen, Denmark
\\
$\ast$ E-mail: Corresponding andersx@nano.ku.dk
\end{flushleft}

% Please keep the abstract between 250 and 300 words
\section*{Abstract}

We present the ProCS method for the rapid and accurate prediction of protein backbone amide proton chemical shifts - sensitive probes of the geometry of key hydrogen bonds that determine protein structure.
ProCS is parameterized against quantum mechanical (QM) calculations and reproduces high level QM results obtained for a small protein with an RMSD of 0.25 ppm (\textit{r} = 0.94).
ProCS is interfaced with the PHAISTOS protein simulation program and is used to infer statistical protein ensembles that reflect experimentally measured amide proton chemical shift values.
Such chemical shift-based structural refinements, starting from high-resolution X-ray structures of Protein G, ubiquitin, and SMN Tudor Domain, result in average chemical shifts, hydrogen bond geometries, and trans-hydrogen bond ($^{\mathrm{h3}}$\textit{J}$_{\mathrm{NC'}}$) spin-spin coupling constants that are in excellent agreement with experiment.
We show that the structural sensitivity of the QM-based amide proton chemical shift predictions is needed to obtain this agreement.
The ProCS method thus offers a powerful new tool for refining the structures of  hydrogen bonding networks to high accuracy with many potential applications such as protein flexibility in ligand binding.

\section*{Introduction}
Chemical shifts hold valuable structural information that is being used increasingly in the determination of protein structure and dynamics\cite{Filatov}.  This is made possible primarily by empirical chemical shift predictors such as SHIFTS, SPARTA, SHIFTX, PROSHIFT, and CamShift \cite{MoonCase2007,XuCase2001,sparta,SHIFTX,PROSHIFT,CamShift}.
While these methods generally offer quite accurate predictions, the predicted chemical shifts of backbone amide protons ($\delta_{\mathrm{H}}$) tend to be significantly less accurate than, for example, the proton on the $\alpha$-carbon \cite{WishartCase,Case2013}.
This is unfortunate since $^{15}$N-HSQC forms a large fraction of all protein NMR studies and $\delta_{\mathrm{H}}$ holds valuable information about the hydrogen bond geometry of the ubiquitous amide-amide hydrogen bonds that are key to protein secondary structure.
Parker, Houk and Jensen\cite{Parker} have proposed a $\delta_{\mathrm{H}}$-predictor that was shown to offer significantly more accurate predictions, although this was only demonstrated for 13 $\delta_{\mathrm{H}}$-values.
The method suggests that there is an exponential dependence of $\delta_{\mathrm{H}}$ in the NH$\cdot\cdot$O=C bond length (as suggested by Barfield\cite{Barfield} and Cornilescu \textit{et al.}\cite{Cornilescu}) as well as a non-negligible contribution from cooperative effects in hydrogen bonding networks.
This exponential dependence makes empirical parameterizations of $\delta_{\mathrm{H}}$-predictors challenging since even small discrepancies between the structure used in the parameterization (usually an X-ray structure without explicitly represented hydrogens) and the solution-phase structural ensemble that gives rise to the experimentally observed $\delta_{\mathrm{H}}$-values can have a significant effect.
The method by Parker \textit{et al.} addresses this problem by parameterization against $\delta_{\mathrm{H}}$-values obtained by quantum mechanical (QM) calculations, and is similar in spirit to the QM-based $\alpha$-carbon chemical shift predictor CheShift developed by Vila \textit{et al.}\cite{CheShift1,CheShift2}.
Both studies noted that the QM-based chemical shift predictors tend to be more sensitive to small structural changes compared to popular empirical chemical shift predictors and therefore promises to be valuable tools in protein structure validation and refinement.
Here we present several key advances in the use of backbone amide proton chemical shifts to refine and validate the geometry of the amide-amide hydrogen bonding network in proteins.
First we present and validate the ProCS method which extends the QM-based backbone amide proton chemical shift predictor proposed by Parker \textit{et al.}\cite{Parker}.
Second we present a computational methodology for using ProCS and experimental $\delta_{\mathrm{H}}$-values to refine the hydrogen bond-geometries of proteins.
This is accomplished by implementing ProCS in the Markov chain Monte Carlo (MCMC) protein simulation framework PHAISTOS\cite{Phaistos}, and using this in combination with a molecular mechanics (MM) force field.
Third, we show for a number of small proteins that structural refinement against experimental $\delta_{\mathrm{H}}$ values using ProCS leads to hydrogen bond geometries that are in closer agreement with high-resolution X-ray structures and experimental trans-hydrogen bond spin-spin coupling constants ($^{\mathrm{h3}}J_{\mathrm{NC'}}$) compared to using an energy function based on the empirical chemical shift predictor CamShift \cite{CamShift} or solely using a force field (OPLS-AA/L\cite{OPLS} with the GB/SA continuum solvent model\cite{gbsa}).

\section*{Results and discussion}
\subsection*{The ProCS method} The ProCS program uses a modified implementation of the formula developed by Parker \textit{et al.}\cite{Parker} where the amide proton chemical shift is approximated by a sum of additive terms:
\begin{equation}
\label{eq:additive_equation}
\delta_{\mathrm{H}} = \delta_{\mathrm{BB}} + \Delta\delta_{\mathrm{1^\circ HB}} + \Delta\delta_{\mathrm{2^\circ HB}} + \Delta\delta_{\mathrm{3^\circ HB}}  +  \Delta\delta_{\mathrm{RC}}
\end{equation}
Here, $\delta_{\mathrm{BB}}$ is a backbone term that depends on the $(\phi, \psi)$ torsion angles of the residue, $\Delta\delta_{\mathrm{1^\circ HB}}$ is due to a primary hydrogen bond directly to the amide proton in question, $\Delta\delta_{\mathrm{2^\circ HB}}$ is due to a secondary hydrogen bond to the carbonyl oxygen in the amide group, $\Delta\delta_{\mathrm{3^\circ HB}}$ is a small term that incorporates further polarization due to hydrogen bonding at the primary and/or secondary bonding partner and $\Delta\delta_{\mathrm{RC}}$ describes magnetic perturbations due to ring currents in nearby aromatic side chains. 
ProCS calculates amide proton chemical shift values referenced to dimethyl-silapentane-sulfonate (DSS).

We have replaced the original $\delta_{\mathrm{BB}}$ term, which was a crude 3-step function, by a scaled version of the $(\phi, \psi)$ backbone torsion angle hypersurface parametrized by Czinki and Cs\'{a}sz\'{a}r\cite{Czinki2004}. The $\delta_{\mathrm{BB}}$ term is given as
\begin{equation}
\delta_{\mathrm{BB}}= 0.828 \cdot \left(\mathrm{ICS}(\phi, \psi) + 0.77 \mathrm{\ ppm} \right)
\end{equation}
where $\mathrm{ICS}(\phi, \psi)$ is the $n$-th order cosine series given in reference \cite{Czinki2004}. The scaling is necessary to account for differences in choice of basis set and molecular geometry optimization\cite{Rablen}.

In the cases described by Parker \textit{et al.}, $\Delta\delta_{\mathrm{RC}}$-values are obtained through the SHIFTS web-interface\cite{XuCase2001}. Since this would be impractical, we implemented the point-dipole\cite{Pople1,Pople2} approximation given by:
\begin{equation}
\Delta\delta_{\mathrm{RC}} = i\ B\ \frac{1-3 \cos^2(\theta)}{|\vec{r}|^3}
\end{equation}
where $i$ is an intensity parameter which depends on the type of aromatic ring, $B$ is a constant of 30.42 ppm \AA$^3$, $\vec{r}$ is the vector between the amide proton and the center of the aromatic ring and $\theta$ is the angle between $\vec{r}$ and the normal to the plane of the aromatic ring located on its center. The values of $i$ and $B$ are obtained from the parameter set by Christensen \textit{et al.}\cite{Christensen2011}. 

The following expression for $\Delta\delta_{\mathrm{1^\circ HB}}$ was implemented for primary bonds to backbone amide carbonyl oxygen atoms:

\begin{eqnarray}
\label{eq:Barfield}
  &\Delta\delta_{\mathrm{1^\circ HB}} = \left[\right. 4.81\cos^2(\theta) + \sin^2(\theta)\left\{\right. 3.10 \cos^2(\rho)\nonumber\\
  & -0.84 \cos(\rho) + 1.75 \left. \right\}\left.\right] e^{-2.0\ \textrm{\AA}^{-1}(r_{\mathrm{OH}} - 1.760\ \textrm{\AA})} \cdot 1\ \textrm{ppm}
\end{eqnarray}

This formula originates from the works of Barfield\cite{Barfield} and is fitted to chemical shifts computed for model systems of hydrogen bonding between two formamide molecules. 
In order to treat hydrogen bonding to other oxygen atom types (carboxylic acids and alcohols as found in side chains and C-terminal), we carried out similar scans (see Supplementary Information, section S2 and Fig. S4) over bond angles and lengths and stored these in lookup-tables from which the chemical shift perturbation due to any hydrogen bonding geometry can be interpolated. 
Hydrogen bonding to carboxylic acid oxygen atoms interaction were modeled by $N$-methylacetamide/acetate dimers, while bonds to alcohols oxygen atoms were modeled by $N$-methylacetamide/methanol dimers.

For non-hydrogen bonding amide protons, which are found primarily on the protein surface, $\Delta\delta_{\mathrm{1^\circ HB}}$ is approximated as the interaction between a water molecule and an $N$-methylacetamide molecule.
In this case, $\Delta\delta_{\mathrm{1^\circ HB}}$ is equal to 2.07 ppm for an energy minimized bonding geometry (see Supplementary Information, section S3and Fig. S5). 
The functional forms of $\Delta\delta_{\mathrm{2^\circ HB}}$ and $\Delta\delta_{\mathrm{3^\circ HB}}$ were kept as described in reference \cite{Parker}.

\subsection*{Reproducing QM chemical shifts}
ProCS predictions result from several terms [Eq.~\ref{eq:additive_equation}] that are assumed to be additive. 
To test this additivity assumption we use density functional theory (DFT) and compute chemical shielding values (at the B3LYP/cc-pVTZ/PCM level) for the crystal structure of human parathyroid hormone, residues 1-34 at 0.9 \AA\ resolution, PDB-code 1ET1\cite{1et1}. Chemical shift values for amide protons at the termini are excluded from the statistics presented in this section, since they do not participate in any hydrogen bonds in the crystal structure.
Using the linear scaling method due to Jain \textit{et al.}\cite{Jain} similar DFT calculations reproduce experimental proton chemical shifts of a test set of 80 small to medium sized molecules to an RMSD of 0.13 ppm.\cite{Jain}

ProCS reproduces the QM calculation with an RMSD of 0.25 ppm (Table \ref{tab:qm}) based on the same structure.  ProCS is parameterized based on a number of DFT calculations (see Methods section) which have been shown to yield proton chemical shifts within 0.16 ppm of experimental values for small organic molecules\cite{Rablen}. Thus, the error from non-additivity is roughly the same as the expected deviation from experiment. 

\begin{figure}[!ht]
\begin{center}
    \includegraphics[width=\textwidth]{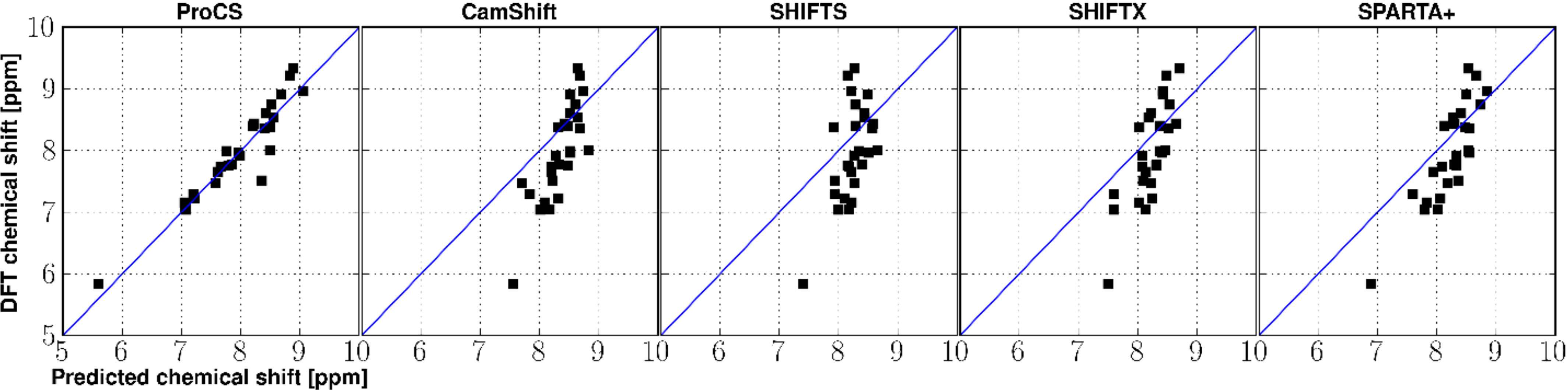}
    \caption{Correlation between chemical shift predictions from five different NMR prediction methods and quantum mechanical chemical shifts for human parathyroid hormone, residues 1-37 (PDB code: 1ET1). Blue lines represent a 1-to-1 correlation.}
    \label{fig:qm_1et1}
\end{center}
\end{figure}

The chemical shifts predicted by empirical methods do not agree well with the DFT results, with RMSD values ranging from 0.56 to 0.70 ppm (see Table \ref{tab:qm} and Fig.~\ref{fig:qm_1et1}).
The DFT chemical shifts span a relatively large range (5.8 - 9.3 ppm) while the empirically predicted chemical shifts span a very narrow range (up to 6.9 - 8.9 ppm for SPARTA+) - see Fig.~\ref{fig:qm_1et1}.  
This indicates that the empirical methods are less sensitive to small differences in hydrogen bond geometry found in the X-ray structure.

\subsection*{Reproducing experimental chemical shifts from X-ray structures}
The QM method used here reproduces small molecule $^1$H chemical shifts with an RMSD of 0.13 ppm\cite{Jain}.  The RMSD between the chemical shifts calculated by QM using the static X-Ray structure and the experimental data obtained in solution is 0.66 ppm. The main sources of this discrepancy are likely inaccuracies in the hydrogen bond lengths in the X-ray structure compared to solution, since there is an exponential dependence of the proton chemical shifts on this distance [Eq.~\ref{eq:Barfield}], and/or the use of a single structure rather than a structural ensemble.

The corresponding RMSD to experimental data for ProCS (0.63 ppm) is similar to the QM RMSD and significantly larger than the 0.25 ppm RMSD between QM and ProCS, indicating that ProCS is sufficiently accurate to identify inaccuracies in the X-ray structure, and/or the effect of using a single structure rather than a structural ensemble. A similar comparison to experiment for 13 other proteins is given in Table \ref{tab:comp} (PDB-codes: 1BRF, 1CEX, 1CY5, 1ET1, 1I27, 1IFC, 1IGD, 1OGW, 1PLC, 1RGE, 1RUV, 3LZT, 5PTI).
The deviation from experiment for the empirical methods are significantly smaller than for ProCS with RMSD values ranging from 0.46 to 0.64 ppm (Table \ref{tab:comp}).
A likely explanation for this is that the empirical methods are parameterized using X-ray structures. In order for these methods to produce low RMSD values relative to experiment they need to be insensitive to errors in protein structure.

\subsection*{Refining protein structures based on chemical shifts}
If indeed the difference in experimental and computed chemical shifts reports on inaccuracies in the protein structure, then minimizing this difference can be used for structural refinement. 
To test this hypothesis we generate structural ensembles that minimizes the difference in computed and observed chemical shifts to the specified uncertainty in the chemical shift model and determine the quality of these structures by comparison to experimental structures and coupling constants (next section).

Refinement is accomplished using a Markov chain Monte Carlo (MCMC) technique described in detail in the Methods section. In short, the method involves Monte Carlo sampling of structural changes using a posterior distribution constructed using the OPLS-AA/L fore field\cite{OPLS} with the GB/SA implicit solvent model\cite{gbsa} (referred to hereafter simply as "OPLS") and amide proton chemical shifts differences from experiment computed using either CamShift or ProCS. We note that the resulting ensemble is not a dynamic ensemble but an ensemble that reflects experimentally measured amide proton chemical shifts. The simulation lengths are roughly equivalent to 6-10 ns of molecular dynamics simulations \cite{crisp}.
We refine the structure of ubiquitin, Protein G, and SMN Tudor domain each based on three energy functions: OPLS alone, OPLS+ProCS and OPLS+CamShift.
Each MC refinement results in an ensemble of 24,000 structural samples for Ubiquitin and 40,000 for Protein G and SMN Tudor Domain, from which average chemical shifts for each amide proton are computed.  The results are summarized in Table \ref{tab:ensembles}. 

The average ProCS chemical shifts are in better agreement with experiment (RMSD 0.81 ppm) compared to using X-ray structures (RMSD 1.10 ppm). The respective RMSD values for amide protons hydrogen bonded to backbone amide groups, other hydrogen bonds, and no hydrogens bonds are 0.31 ppm, 0.78 ppm and 1.09, respectively.  These RMSD values reflect the uncertainties defined for each kind of hydrogen bonding situation in the ProCS model (see Methods section) meaning that the simulations have indeed converged to a distribution of structures reflecting the experimental chemical shifts within the accuracy of the ProCS model at the given temperature. 
A corresponding structural ensemble generated solely from the OPLS force field increases the RMSD from experiment to 1.52 ppm, indicating more inaccurate hydrogen bond geometries (more on this in the next section).

An MC-based structural refinement based on OPLS and chemical shifts derived from CamShift has no substantial effect on the chemical shift RMSD compared to the X-ray structure (0.50 vs 0.46 ppm). Using the OPLS-derived structural ensemble increases the RMSD by 0.1 ppm compared to using X-ray structures when CamShift is used to calculate chemical shifts. This indicates that an OPLS-based refinement does not improve the hydrogen bonding geometry and that CamShift is less sensitive to a change in structure compared to ProCS.

\subsection*{Hydrogen bond geometries}
\begin{figure}[!ht]
\centerline{\includegraphics[width=8.7cm]{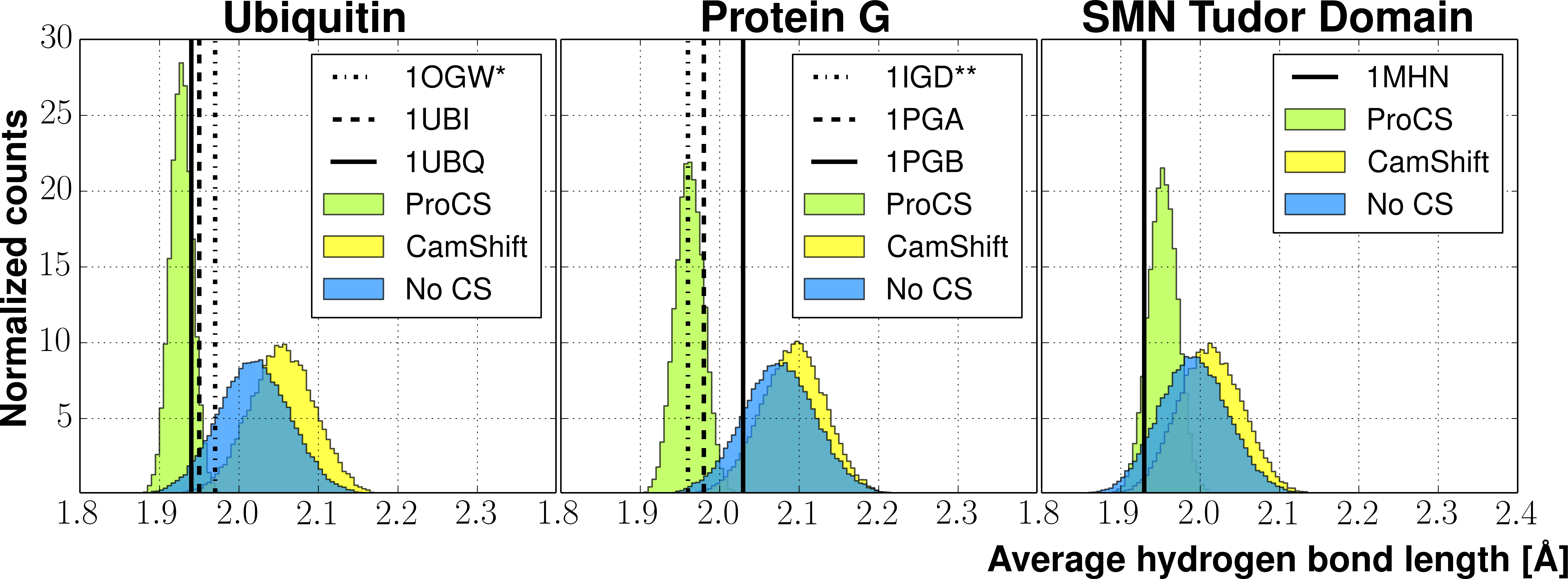}}
\caption{Distribution of average hydrogen bond lengths throughout Monte Carlo simulations on Ubiquitin, Protein G and SMN Tudor Domain. Histograms are normalized (to an area of 1) to fit identical axes.
Vertical lines indicate average values obtained from experimental X-ray structures (PDB-codes are noted in the figure legends).
The blue histogram represents the simulation with only the molecular mechanics energy from the OPLS-AA/L force field with the GB/SA solvent model (but no chemical shift energy term). Green and yellow histograms indicate the use of OPLS force field plus an additional chemical shift energy term from ProCS or CamShift, respectively. *1OGW contains fluoro leucine at residues 50 and 67. **1IGD is a closely related homologue (see text).}
\label{fig:hbond_dist}
\end{figure}

\begin{figure}[!ht]
%\centerline{\includegraphics[width=8.7cm]{3_2d_corrected.tiff}}
\centerline{\includegraphics[width=8.7cm]{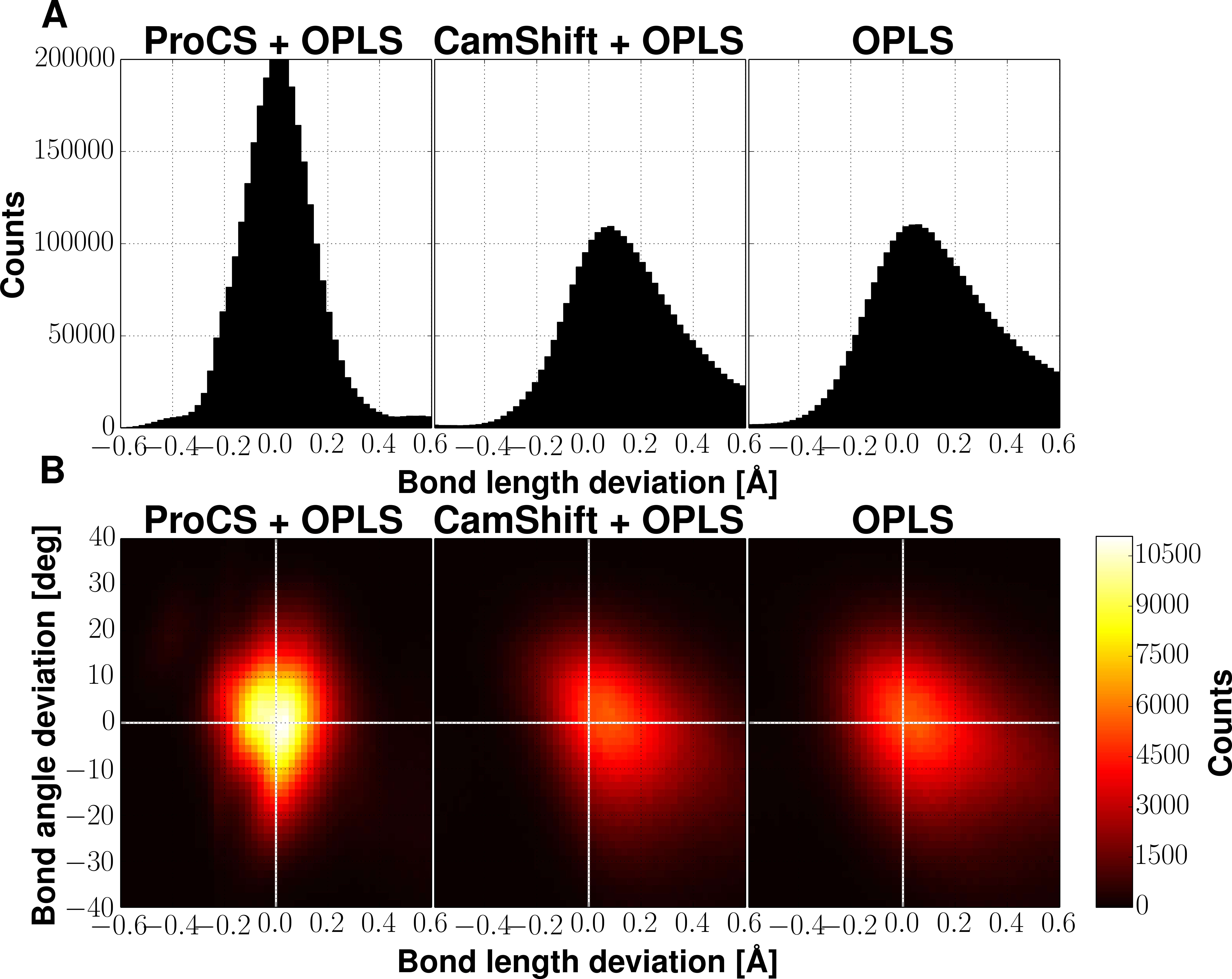}}
\caption{Deviation in hydrogen bonding geometries between the experimental X-ray structure and samples obtained from Markov Chain Monte Carlo (MCMC) simulations using the OPLS-AA/L force field with the GB/SA solvent model with either no chemical shift energy term or a chemical shift energy from either ProCS or CamShift. Data is calculated over all amide-amide bonding pairs for which experimental $^{\mathrm{h3}}J_{\mathrm{NC'}}$ spin-spin coupling constants were present. (A) shows the distribution of the deviations found in the MCMC ensembles from the experimental hydrogen bond length found in the X-ray structure. (B) shows the correlation of deviations in hydrogen bond lengths and H$\cdot \cdot$O=C bond angles from the experimental X-ray structures.}
\label{fig:hbond_geometries}
\end{figure}

The H$\cdot \cdot$O distances and H$\cdot \cdot$O=C angles of the backbone amide-amide hydrogen bonds for which $^{\mathrm{h3}}J_{\mathrm{NC'}}$ coupling constants have been measured (see next section) are extracted from the ensembles and compared to the corresponding values found in the experimental X-ray structures with hydrogens added from PDB2PQR\cite{pdb2pqr1,pdb2pqr2}.  The result are shown in Table \ref{tab:ensembles} and Figures \ref{fig:hbond_dist} and \ref{fig:hbond_geometries}.

Fig.~\ref{fig:hbond_dist} shows the distributions of H$\cdot \cdot$O distances from the ensembles computed using the three energy terms described in the previous section. Structural refinement using OPLS and ProCS for ubiquitin results in ensembles with average H$\cdot \cdot$O distances that have an RMSD within 0.02 \AA\ of those found in the X-ray structures 1UBQ and 1UBI (both 1.80 \AA\ X-ray resolution) and 0.04 \AA\ from the ubiquitin structure 1OGW (1.30 \AA\ X-ray resolution) in which the leucine residues 50 and 67 have been replaced by fluoro leucine. For Protein G we note that the resulting ensemble does not have an average H$\cdot \cdot$O distance that agrees well (0.07 \AA\ difference) with the starting structure 1PGB (1.92 \AA\ X-ray resolution). However the difference from the 1PGA structure (2.07 \AA\ X-ray resolution) and the more accurate 1IGD structure (X-ray resolution of 1.1 \AA) is much less, 0.02 \AA\ and 0.00 \AA, respectively. The 1IGD structure is a close homologue which has 89\% sequence identity score and 95\% sequence similarity.
In the case of the SMN Tudor Domain, ProCS-based refinement results in slightly longer amide-amide hydrogen bond lengths (0.02 \AA\ on average) compared to the X-ray structure 1MHN.

In contrast, structural refinement using CamShift and OPLS or just OPLS leads to increases in average H$\cdot \cdot$O bond lengths of up to 0.15 \AA, with a standard deviation 2-3 times larger than that found in the OPLS+ProCS simulation. 
In all cases use of CamShift has relatively little effect on the ensemble average H$\cdot \cdot$O distance compared to just using OPLS.

In all cases, the use of ProCS leads to a significantly smaller standard deviation in H$\cdot \cdot$O bond lengths: 0.017 \AA\ compared to 0.045 and 0.041 \AA\ for CamShift+OPLS and OPLS, respectively (Fig.~\ref{fig:hbond_geometries}A).
The H$\cdot \cdot$O=C bond angles observed in the ProCS+OPLS simulations are on average within $-2.0^\circ$ of corresponding value observed in the X-ray structures. The same bond angle differences are $-6.7^\circ$ and $-7.4^\circ$ observed in the CamShift+OPLS and OPLS simulations, respectively (Fig.~\ref{fig:hbond_geometries}B).

\subsection*{Trans-hydrogen bond coupling constants}
\begin{figure}[!ht]
%\centerline{\includegraphics[width=8.7cm]{4_correl_j_1UBQ_corrected.tiff}}
\centerline{\includegraphics[width=8.7cm]{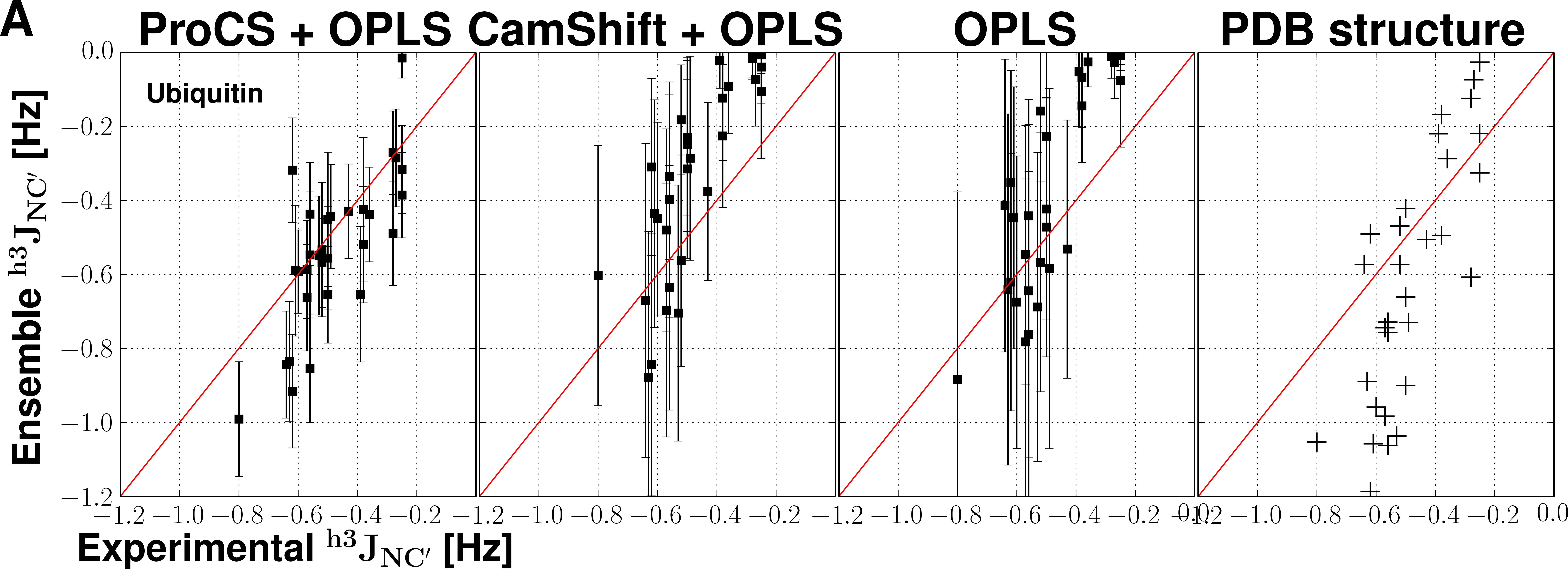}}
%\centerline{\includegraphics[width=8.7cm]{5_correl_j_1MHN.tiff}}
\centerline{\includegraphics[width=8.7cm]{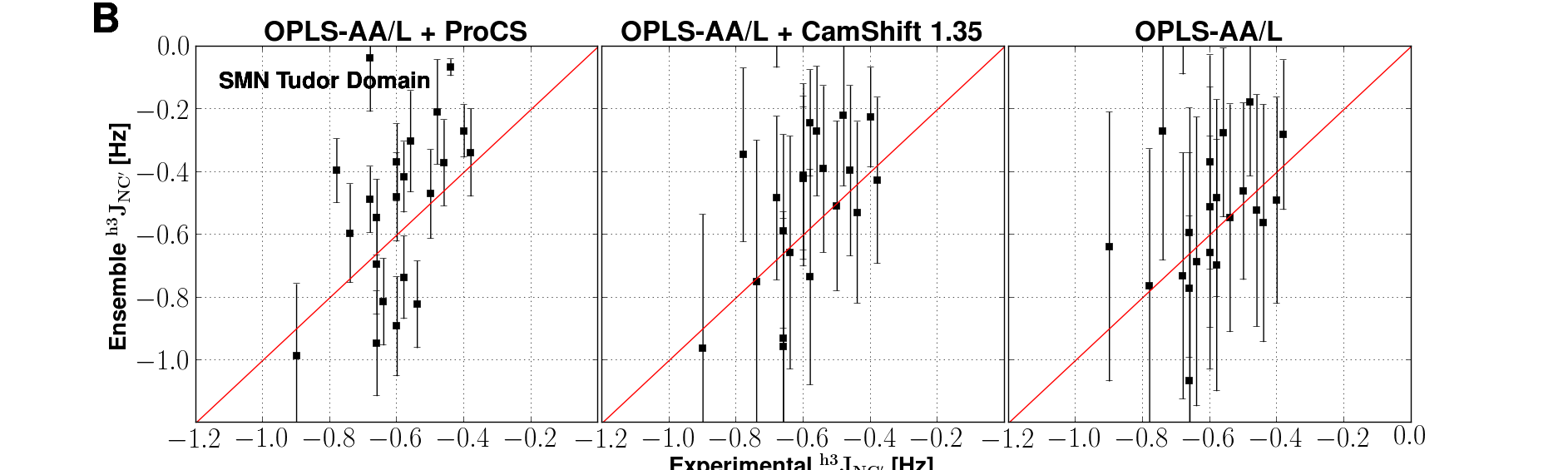}}
%\centerline{\includegraphics[width=8.7cm]{6_correl_j_1PGB_corrected.tiff}}
\centerline{\includegraphics[width=8.7cm]{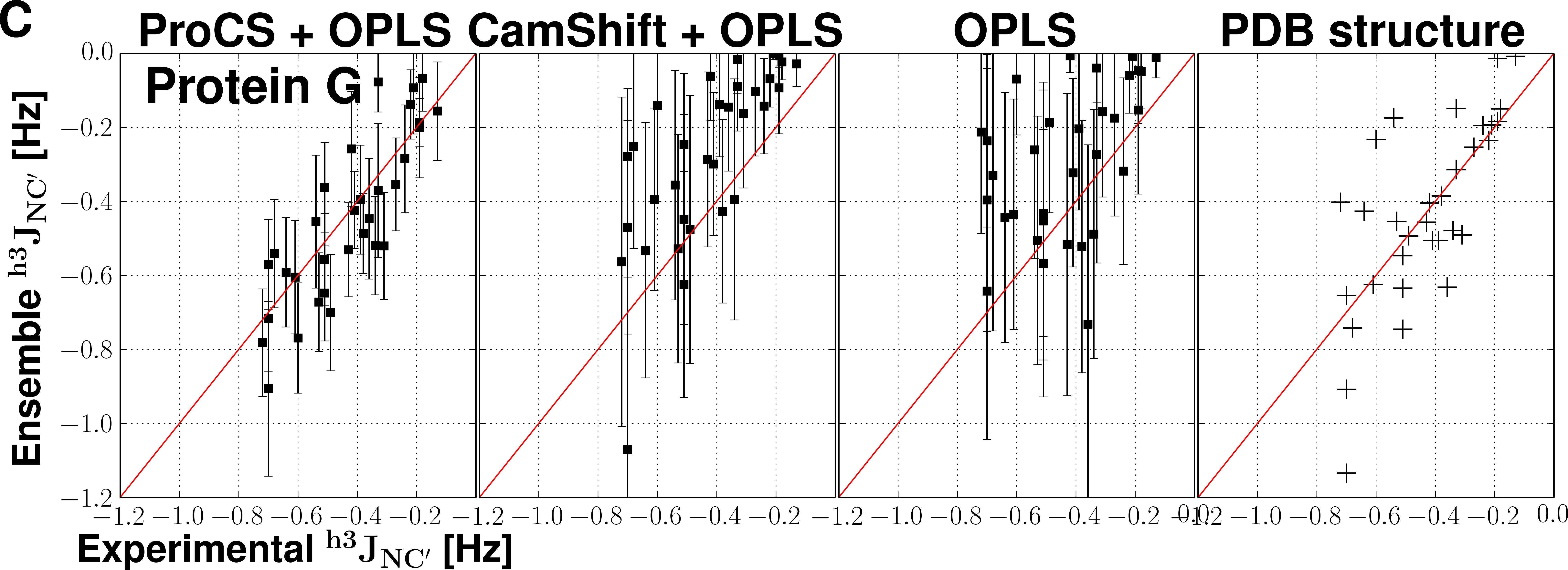}}
\caption{Reproducing experimental $^{\mathrm{h3}}J_{\mathrm{NC'}}$ spin-spin coupling constants via different structural ensembles and experimental X-ray structures.
Squares denote the average coupling constant observed for that hydrogen bond in the ensemble and error bars represent the standard deviation observed throughout the simulations. Crosses represent the spin-spin coupling constants calculated using the static experimental X-ray structure.
Results from simulations on ubiquitin is displayed in A, SMN Tudor domain in B and Protein G in C.
Left column displays simulations only the OPLS-AA/L force field with the GB/SA solvent model (OPLS) and the ProCS energy term; second column is from OPLS plus the CamShift energy term; thrid column is for the simulation with only the OPLS force field energy. In the rightmost column $^{\mathrm{h3}}J_{\mathrm{NC'}}$ are computed from the corresponding X-ray structure.}
\label{fig:J-coupling}
\end{figure}

Better agreement with X-ray structures does not necessarily imply better solution-phase structures. In order to compare the resulting ensembles to solution-phase data we compute average trans-hydrogen bond coupling constants and compare these to experimental values.
Experimental trans-hydrogen bond $^{\mathrm{h3}}J_{\mathrm{NC'}}$ spin-spin coupling constants represent a very sensitive measure for solution-phase hydrogen bonding conformations and are known to correlate with amide proton chemical shifts\cite{Cordier}.
The coupling constants depend exponentially on the hydrogen bonding distance and on bond angles\cite{Barfield}.
Data from ensemble back-calculated $^{\mathrm{h3}}J_{\mathrm{NC'}}$ spin-spin coupling constants are summarized in Fig.~\ref{fig:J-coupling} and Table \ref{tab:ensembles}.

In the ubiquitin simulations, the OPLS force field on its own does not yield ensemble $^{\mathrm{h3}}J_{\mathrm{NC'}}$ averages in good agreement with experimental data. 
In this simulation, several hydrogen bonds were eventually broken. Calculated $^{\mathrm{h3}}J_{\mathrm{NC'}}$-values for these partly unfolded hydrogen bonds show up close to 0 Hz (see Fig.~\ref{fig:J-coupling}A). The RMSD to experimental values is here 0.18 Hz. Adding the energy term from amide proton chemical shifts via CamShift does not help keeping these hydrogen bonds fixed, but results in a minor improvement in RMSD to 0.17 Hz.
Adding the amide proton chemical shifts energy term via ProCS to the OPLS force field stabilized the hydrogen bonds and also gave an improvement in the RMSD values to 0.14 Hz, which is close to that of the most accurate structural NMR ensembles of ubiquitin (see Table \ref{tab:ubiquitin}). For Protein G we obtained similar RMSD values: 0.20 Hz, 0.14 Hz and 0.18 Hz for the OPLS alone, OPLS+ProCS and the OPLS+CamShift simulations, respectively.
In the SMN Tudor Domain simulation, the average $^{\mathrm{h3}}J_{\mathrm{NC'}}$ value of all three types of simulations were comparably close to experimental values 0.24, 0.24 and 0.23 Hz for OPLS alone, OPLS+ProCS and the OPLS+CamShift simulations, respectively.
Thus, overall the coupling constants based on the ProCS refined ensembles are indeed in better agreement with experimental values indicating the refinement led to improved hydrogen bond geometries compared to using OPLS or OPLS+CamShift.

\subsection*{Impact on Q-factor}

In this section we investigate how amide proton chemical shifts restraints affect back-calculated $^1D_{NH}$ residual dipolar couplings (RDCs) compared to experimental values for ubiquitin. RDCs are attractive in this regard since they report on structural features that are not related to hydrogen bonding conformations as studied intensively in the previous sections.
The Q-factor is a qualitative measure for the agreement between back-calculated RDCs and the corresponding experimentally observed values\cite{qfactor}.

We find, that for our Ubiquitin ensemble generated using the OPLS force field alone has a Q-factor of 0.29 while inclusion of chemical shifts only gives a very modest improvement of this figure to 0.27 for both CamShift and ProCS as chemical shift model.
The same value calculated for the three X-ray structures 1UBQ, 1UBI and 1OGW are 0.22, 0.25 and 0.26, respectively. For six NMR-based ensembles the Q-factor is in the range 0.04-0.38, though in some cases the ensembles were refined against the RDCs (see Table \ref{tab:ubiquitin}). We observe no significant correlation ($P < 0.05$) between RMSDs for predicted chemical shifts or spin-spin couplings constant to their experimental values and the calculated Q-factor for the 12 cases presented in Table \ref{tab:ubiquitin}.

While amide proton chemical shifts have some dependence on the dihedral angles of the backbone, the dependence on the particular hydrogen bonding conformations is much larger in comparison. This is due to an exponential dependence on the hydrogen bond length. 

The distribution from which we sample chemical shifts is constructed from a prior distribution based on the OPLS force field and a likelihood which contains information from experimental chemical shifts. We expect that structural features of the resulting ensemble, which are not local to the hydrogen bond geometry, will largely reflect the prior distribution, i.e. in this our case, the OPLS force field. 

\subsection*{Computational efficiency}
Executing the simulations on one core of a Intel Xeon X5560 running at 2.80 GHz with the 1UBQ structure, the average evaluation time of the three different energy-terms were OPLS-AA/L: 27ms, CamShift 1.35: 4.7ms, ProCS: 0.74ms. Similar evaluation times were observed for the 1MHN and 1PGB simulations. Note that, in our implementation, the CamShift term calculates chemical shifts for six atoms per residue, even if those chemical shifts are not a used to evaluate the corresponding energy term.
The OPLS and CamShift terms were implemented with a caching algorithm, so only the subset of parts of the chemical shift terms that change after a local Monte Carlo move were recomputed.
This approach was not implemented for ProCS since the OPLS force field energy evaluation is by far the most computationally expensive step.
Running on four cores, we obtained between 10 to 16 mio Monte Carlo iteration steps total \textit{per} day, depending on the protein size and combination of energy terms.

\section*{Methods}
\subsection*{Monte Carlo refinement of protein structure}

We employ Markov chain Monte Carlo sampling from a Bayesian posterior distribution to perform protein structure refinements and simulations. 
MCMC simulations are attractive because no gradient expressions need to be derived for ProCS. 
Bayesian inference\cite{Rieping2005} provides a rigorous mathematical framework for the inference of protein structure from experimental data.
It involves the construction of a posterior distribution, which consists of a prior distribution and a likelihood. 
The former brings in general information on protein structure, and in our case is based on the OPLS energy function. 
The latter brings in the experimental data, and is based on the difference between the back-calculated data from a simulated structure and the experimental data.
Using PHAISTOS, we draw samples from the joint probability distribution, which is given by:
\begin{equation}
p\left(X|\left\{\delta_i^{\mathrm{exp}}\right\},I\right) \propto p\left(\left\{\delta_i^{\mathrm{exp}}\right\}|X,I\right) p\left(X|I\right)\label{eqP}
\end{equation}
where $X$ represents a protein structure, $\left\{\delta_i^{\mathrm{exp}}\right\}$ is experimental chemical shift data and $I$ denotes prior information, such as sequence and knowledge about the uncertainties in the prediction model. The prior distribution $p\left(X|I\right)$ is proportional to $\exp\left(-\beta E_{\mathrm{FF}}\right)$, where $E_{\mathrm{FF}}$ is the molecular mechanics force field potential energy and $\beta = \slfrac{1}{k_\mathrm{B}T}$. $p\left(\left\{\delta_i^{\mathrm{exp}}\right\}|X,I\right)$ denotes the probability of observing experimental data given a trial structure.
Under the assumption that the error in the chemical shift prediction model follows a Gaussian distribution with some set of standard deviations $\{\sigma_i\}$, the expression for $p\left(\left\{\delta_i^{\mathrm{exp}}\right\}|X,I\right)$ is:
\begin{equation}
\label{eqn:padawan-energy}
p\left( \left\{\delta_i^{\mathrm{exp}}\right\}|X,\left\{\sigma_{i} \right\}\right) = \prod_{i=1}^{n}\left[ \sqrt{\frac{1}{2 \pi\sigma_i^2 }}\exp\left\{-\frac{\left(\Delta \delta_i\right)^2}{2\sigma_i^2}\right\}\right]
\end{equation}
where $\Delta \delta_i$ is the discrepancy between predicted and experimental data for the $i$-th nucleus of the data set in the trial structure, $X$.
This formulation of the posterior distribution assumes that the prior distribution on X is also a good prior distribution for the chemical shift differences, $\Delta \delta_i$, otherwise an additional term would be required\cite{pmf}.
The set of standard deviations, $\left\{\sigma_i\right\}$ was assigned based on the primary bond type, since, for instance, the model for solvent exposed amide protons is much cruder than the amide-amide bonding model. $\sigma_i$ was set to 0.3 ppm, for primary bonds to another backbone amide, 0.5 ppm to a side chain amide group, 0.8 ppm to a side chain alcohol or carboxylic acid group and 1.2 ppm for solvent exposed amide protons and other types of bond not included in the prediction model.

\subsection*{Protein Structures and NMR data}
All protein structures used in this study were downloaded from the RCSB Protein Data Bank\cite{PDB} (PDB) and protonated using PDB2PQR 1.5, \cite{pdb2pqr1,pdb2pqr2} with PROPKA\cite{propka} to determine protonation states at the pH at which NMR data was recorded. Chemical shift data were obtained from the RefDB\cite{refdb} or the Biological Magnetic Resonance Bank\cite{bmrb}, and subsequently re-referenced through Shiftcor\cite{refdb}.  $^{\mathrm{h3}}J_{\mathrm{NC'}}$ spin-spin coupling constants for 1PGB, 1UBQ and 1MHN were obtained from references \cite{Cordier}, \cite{Cornilescu} and \cite{Markwick}, respectively. %The estimated experimental error on $^{\mathrm{h3}}J_{\mathrm{NC'}}$ values is 0.02 Hz.

\subsection*{MCMC simulations} MCMC simulations were carried out in PHAISTOS v1.0-rc1 (rev. 335) using the Metropolis-Hastings algorithm at 300 K. The simulations are initialized from the experimental crystal structures.
Four independent trajectories were simulated for each protein structure.
A total of 100 mio MC steps were taken for each trajectory for Protein G and the SMN Tudor Domain simulation and 85 mio MC steps for the Ubiquitin simulation.
Structures were saved every 10,000 Monte Carlo step.
The Monte Carlo move-set was composed of 25\% CRISP backbone moves\cite{crisp} and 75\% uniform side chain moves.
The force field energy was calculated using the OPLS-AA/L force field \cite{OPLS} with the GB/SA continuum solvent model \cite{gbsa}. The following crystal structures obtained from the PDB were used as starting structures in the simulations: 1PGB (Protein G), 1UBQ (Ubiquitin) and 1MHN (SMN Tudor Domain).
Time evolution of Monte Carlo energy and chemical shift RMSDs are available in the Supplementary Information (section S1, figures S1-S3).

\subsection*{Back calculation of spin-spin coupling constants}
$^{\mathrm{h3}}J_{\mathrm{NC'}}$ spin-spin coupling constants were calculated using the approximation by Barfield\cite{Barfield}.
\begin{eqnarray}
&^{\mathrm{h3}}J_{\mathrm{NC'}}(\theta, \rho,r_{\mathrm{OH}})  = \left[\right. - 1.31\cos^2(\theta) + \left\{ \right. 0.62 \cos^2(\rho) + \nonumber\\\label{eqJ}
&\ 0.92 \cos(\rho) + 0.14 \left. \right\}  \sin^2(\theta)\left.\right]  e^{-3.2\ \textrm{\AA}^{-1}(r_{\mathrm{OH}} - 1.760\ \textrm{\AA})} \cdot 1\ \textrm{Hz}
\end{eqnarray}Here, the coupling depend on the $\angle$N-H$\cdot \cdot$O=C angle, $\rho$,  $\angle$H$\cdot \cdot$O=C,  $\theta$, and the hydrogen bonding distance, $r_{\mathrm{OH}}$. From the MCMC ensembles, the mean $^{\mathrm{h3}}J_{\mathrm{NC'}}$ spin-spin coupling constant was calculated via Eqn.~7 and the standard deviation was calculated as the root mean square deviation from the mean. 
The $^{\mathrm{h3}}J_{\mathrm{NC'}}$ RMSD to experiment is then given as
\begin{eqnarray}
    ^{\mathrm{h3}}J_{\mathrm{NC'}}\ \mathrm{RMSD} &=& \sqrt{\frac{\sum_i \left(^{\mathrm{h3}}J_{\mathrm{NC'}}^{\mathrm{exp},i} - \langle^{\mathrm{h3}}J_{\mathrm{NC'}}^{\mathrm{calc},i} \rangle \right)^2   }{N}}
\end{eqnarray}
where $\langle^{\mathrm{h3}}J_{\mathrm{NC'}}^{\mathrm{calc},i} \rangle$ is the average value over the ensemble for the $i$'th coupling constant.

\subsection*{QM NMR calculations}
All density functional theory (DFT) calculations of NMR isotropic shielding constants involved in the parametrization of ProCS were carried out in Gaussian 03\cite{g03}. Data was obtained at the GIAO/B3LYP/6-311++G(d,p)//B3LYP/6-31+G(d) level of theory using the scaling technique by Rablen \textit{et al.} \cite{Rablen}.
  
The NMR calculation on the 1ET1 protein structure was carried out at the B3LYP/cc-pVTZ/PCM level of theory with a water-like dielectric constant of 78.3553. In this case shielding constants were converted to chemical shifts using the scaling factor obtained by Jain \textit{et al.}\cite{Jain}, assuming that the value of the dielectric constant has a negligible contribution to the scaling factors.

\subsection*{Calculation of ubiquitin Residual Dipolar Couplings}
Residual dipolar couplings were back-calculated from the structural ensembles using singular value decomposition to fit the alignment tensor\cite{rdc}. Ensemble averaging was taken into account so that all structures simultaneously were fitted to a single alignment tensor\cite{rdcensemble}. The agreement to experimental values was calculated via the Q-factor:\cite{qfactor}
\begin{equation}
Q = \frac{\sqrt{\sum \left( \mathrm{RDC}^{\mathrm{exp}} -  \mathrm{RDC}^{\mathrm{calc}} \right)^2 }} { \sqrt{\sum \left( \mathrm{RDC}^{\mathrm{calc}} \right)^2}}
\end{equation}

\section*{Conclusions}
ProCS is a QM-based backbone amide proton chemical shift ($\delta_{\mathrm{H}}$) predictor that can deliver QM quality chemical shift predictions for a protein structure in a millisecond.
$\delta_{\mathrm{H}}$-values predicted using X-ray structures are in worse agreement with experiment, compared to those of the popular empirical chemical shift-predictors CamShift, SHIFTS, SHIFTX,  and SPARTA+.
However the agreement with experiment can be significantly improved by refining the protein structures using an energy function that includes a force field and a solvation term (OPLS-AA/L with the GB/SA continuum solvent model) and a chemical shift term in the program PHAISTOS.
This refinement also results in structures with predicted trans-hydrogen bond coupling constants ($^{\mathrm{h3}}J_{\mathrm{NC'}}$) in good agreement with experiment indicating that the refined protein structures reflect the structures in solution.
Comparison of average hydrogen bond geometries to those of high-resolution ($< 1.35$ \AA) X-ray structures reveals that the structural refinement improves the predicted $\delta_{\mathrm{H}}$-values through relatively small changes in the hydrogen bond geometry distribution.

Structural refinement without chemical shifts (i.e.~using only the OPLS-AA/L + Generalized Born solvation energy) or combined with CamShift has relatively little effect on the predicted $\delta_{\mathrm{H}}$-values, while the predicted $^{\mathrm{h3}}J_{\mathrm{NC'}}$ values are in slightly worse agreement with experiment compared to using X-ray structures or ProCS-refined structures.
This is not surprising given the fact that CamShift and similar empirical methods were designed to be insensitive to relatively small changes in protein structure in order to offer robust chemical shift predictions based on X-ray structures of varying accuracy.
Structural refinement based on other empirical shift predictors, such as SHIFTS, SHIFTX, and SPARTA+, were not tested mainly because an efficient interface to PHAISTOS requires a complete re-implementation of the method.
However, based on our comparison to the QM-calculations (Table \ref{tab:qm} and Fig.~\ref{fig:qm_1et1}) we do not think the conclusions will be substantially different.
Our data, and that of Vila \textit{et al.} \cite{CheShift2}, suggests that QM-derived chemical shift predictors are sufficiently accurate to extract small changes in structure and dynamics from experimentally measured protein chemical shifts.

We are currently working on implementing a QM-based chemical shift prediction method for the remaining H, C, and N nuclei in a protein in ProCS (unfortunately, the source code of the CheShift method developed by Vila \textit{et al.} for QM-based C chemical shift prediction is not available).
The resulting ProCS/PHAISTOS interface should provide a powerful tool for chemical shift-based protein structure refinement.
\\\\The ensembles resulting from the simulations can be downloaded from DOI: http://dx.doi.org/10.5879/BILS/p000001
\\\\Implementations of ProCS and CamShift can be downloaded as separate modules for PHAISTOS under the terms of the GNU General Public License v3 from: http://github.com/jensengroup/

% Do NOT remove this, even if you are not including acknowledgments
\section*{Acknowledgments}

\section*{Supplementary Info}
Section S1: Time evolution of energies and chemical shift RMSDs during MCMC simulation.
Figures S1-S3: Details of Monte Carlo energies and chemical shift RMSDs over time for the presented simulations.
Section S2: Parametrization of chemical shift contributions due to hydrogen bonding interactions to carboxylic acids and alcohols
Figure S4: Sketches showing the geometric parameters and the systems used in the modeling of chemical shift contributions due to hydrogen bonding.
Section S3: Model for solvent exposed amide protons.
Table S4: Chemical shift contributions due to hydrogen bonding to water molecules.
Figure S5: Local minima of NMA-water dimer.

%\section*{References}
% The bibtex filename
\bibliography{procs_plos}

\clearpage

\section*{Figurs}

\section*{Tables}

\begin{table}
%\centering
\begin{threeparttable}
\caption{Correlation coefficients and RMSD between five chemical shift predictors, chemical shifts derived from quantum mechanics (B3LYP/cc-pVTZ/PCM) chemical shifts and experimental values}
\label{tab:qm}
    \begin{tabular}{p{0.11\textwidth} p{0.065\textwidth} p{0.065\textwidth} p{0.065\textwidth}p{0.04\textwidth} c}
      Data source\tnote{a}  & Exp'tl    & Exp'tl  & QM      & QM \\
                                    & $r$       & RMSD    & $r$   & RMSD \\
      \hline
      ProCS                         & 0.54      & 0.63    & 0.94  & 0.25  \\
      SHIFTS\cite{MoonCase2007}     & 0.64      & 0.37    & 0.59  & 0.70  \\
      SHIFTX\cite{SHIFTX}           & 0.69      & 0.37    & 0.71  & 0.62  \\
      SPARTA+\cite{SPARTAPLUS}      & 0.69      & 0.42    & 0.68  & 0.56  \\
      CamShift\cite{CamShift}       & 0.64      & 0.32    & 0.59  & 0.66  %\\
    \end{tabular}
         \begin{tablenotes}
       \item[a] The crystal structure of human parathyroid hormone, residues 1-34 at 0.9 \AA\ resolution (PDB-code 1ET1\cite{1et1}) is used as input structure in all chemical shift calculations.
     \end{tablenotes}
  \end{threeparttable}
    
\end{table}

\begin{table}
\begin{threeparttable}
\caption{Reproduction of experimental amide proton chemical shift values based on 13 X-ray structures with a crystallographic resolution of 1.35 \AA~or less}
\label{tab:comp}

    \begin{tabular}{p{0.17\textwidth} p{0.17\textwidth} l}
        Method & $\langle r \rangle$ \tnote{a} & $\langle$RMSD$\rangle$ \tnote{}\\
        \hline
        ProCS                       &  0.58     &  1.13 ppm  \\
        SHIFTS\cite{MoonCase2007}  &  0.56     &  0.64 ppm  \\
        SHIFTX\cite{SHIFTX}        &  0.71\tnote{c}    &  0.51 ppm\tnote{c} \\
        SPARTA+\cite{SPARTAPLUS}        &  0.79     &  0.40 ppm  \\
        CamShift\cite{CamShift}     &  0.74     &  0.46 ppm
    \end{tabular}
    \begin{tablenotes}
        \item[a] $\langle r \rangle$ denotes the average correlation coefficient over the 13 structure.
        \item[b] $\langle$RMSD$\rangle$ denotes the average root mean square deviation over the 13 structure.
        \item[c] For SHIFTX, three structures displayed over fitting behavior with $r \approx 0.99$. These structures are excluded from the average values.
    \end{tablenotes}
\end{threeparttable}
\end{table}

\begin{table}

\begin{threeparttable}
    \caption{Statistics for three different types of protein simulations}
    \label{tab:ensembles}
        \begin{tabular*}{\hsize}{@{\extracolsep{\fill}}l r r r r}
                                                    & ProCS     & CamShift  &  $\langle$Bond length         & \\
        Structures\tnote{a} & $^1$H RMSD
         &   $^1$H RMSD
          &  deviation$\rangle$\tnote{b}
           &  $^{\mathrm{h3}}J_{\mathrm{NC'}}$ RMSD
            \\
        \hline
        Ubiquitin Ensembles: CamShift + OPLS                   &  0.79 ppm  &  -         & 0.03 \AA  &  0.17 Hz   \\
        Ubiquitin Ensembles: CamShift + OPLS                   & -          &  0.50 ppm  & 0.37 \AA  &  0.17 Hz   \\
        Ubiquitin Ensembles: OPLS (no chemical shifts)         &  1.56 ppm  &  0.60 ppm  & 0.41 \AA  &  0.18 Hz   \\
        1UBQ X-ray starting structure                          &  1.22 ppm  &  0.51 ppm  & -         & 0.22 Hz \\

        SMN Tudor Domain Ensembles: ProCS + OPLS               &  0.93 ppm  & -          & 0.09 \AA  & 0.24 Hz   \\
        SMN Tudor Domain Ensembles: CamShift + OPLS            & -          &  0.46 ppm  & 0.17 \AA  & 0.23 Hz   \\
        SMN Tudor Domain Ensembles: OPLS (no chemical shifts)  &  1.47 ppm  &  0.61 ppm  & 0.22 \AA  & 0.23 Hz   \\
        1MHN X-ray starting structure                          &  1.09 ppm  &  0.65 ppm  & -         & 0.24 Hz \\

        Protein G Ensembles: ProCS + OPLS                      &  0.69 ppm  & -          & 0.06 \AA  &  0.14 Hz   \\
        Protein G Ensembles: CamShift + OPLS                   & -          &  0.52 ppm  & 0.38 \AA  &  0.18 Hz   \\
        Protein G Ensembles: OPLS (no chemical shifts)         &  1.54 ppm  &  0.68 ppm  & 0.37 \AA  &  0.20 Hz   \\
        1PGB X-ray starting structure                          &  1.21 ppm  &  0.55 ppm  & -         &  0.17 Hz
        %\hline 
        %\hline 
    \end{tabular*}
        \begin{tablenotes}
        \item[a] The ensembles are obtained from MCMC simulations using either OPLS-AA/L with the GB/SA solvent model (OPLS) force field energy or OPLS energy plus a chemical shift energy term from from either ProCS or CamShift. Values are calculated over four runs on each of three protein structures, Ubiquitin, Protein G and SMN Tudor Domain, or their static X-ray structure.
        \item[b] The mean bond length deviation denotes the mean absolute difference between the mean hydrogen bond length observed in the sampled structures to the mean hydrogen bond length observed in the corresponding X-ray structure noted below.
    \end{tablenotes}
\end{threeparttable}
\end{table}

\begin{table}

\begin{threeparttable}
    \caption{Statistics for selected ubiquitin ensembles and X-ray structures.\tnote{a}}
    \label{tab:ubiquitin}
    \begin{tabular*}{\hsize}{@{\extracolsep{\fill}}l r r r r r r}
                                  &  (CamShift)  & (CamShift) & (ProCS)& (ProCS)  &  $^{\mathrm{h3}}J_{\mathrm{NC'}}$  &\\
        PDB-ID                    &  $^1$H RMSD  & $r$ & $^1$H RMSD & $r$  &  RMSD   & Q-factor \\\hline
        \tnote{b}2KOX             &  0.29  &  0.84  &  0.68  &  0.86  &  0.12  &  0.04  \\
        \tnote{c}2K39             &  0.34  &  0.82  &  0.98  &  0.77  &  0.13  &  0.07  \\
        \tnote{d}2KN5             &  0.23  &  0.91  &  0.71  &  0.82  &  0.12  &  0.22  \\
        \tnote{e}2NR2             &  0.44  &  0.74  &  1.35  &  0.64  &  0.14  &  0.25  \\
        \tnote{f}1XQQ             &  0.38  &  0.81  &  0.92  &  0.77  &  0.14  &  0.38  \\
        \tnote{g}1D3Z             &  0.41  &  0.79  &  1.00  &  0.71  &  0.30  &  0.06  \\
        \tnote{h}1UBQ             &  0.40  &  0.77  &  0.92  &  0.72  &  0.22  &  0.22  \\
        \tnote{i}1UBI             &  0.40  &  0.77  &  0.97  &  0.73  &  0.33  &  0.25  \\
        \tnote{j}1OGW             &  0.36  &  0.73  &  0.84  &  0.73  &  0.17  &  0.26  \\
        \tnote{k}OPLS + ProCS     &  0.32  &  0.79  &  0.17  &  0.98  &  0.14  &  0.27  \\
        \tnote{k}OPLS + CamShift  &  0.32  &  0.90  &  1.15  &  0.86  &  0.17  &  0.27  \\
        \tnote{k}OPLS             &  0.48  &  0.78  &  1.11  &  0.78  &  0.18  &  0.29 
   \end{tabular*}
   \begin{tablenotes}
        \item[a] Chemical shifts RMSD and $r$ values are calculated for the residues for which $^{\mathrm{h3}}J_{\mathrm{NC'}}$ spin-spin coupling constants have been measured.\cite{Cornilescu}
        \item[b] ERNST method/CHARMM27 + NOE + RDC\cite{2KOX}
        \item[c] OPLS/AA-L + NOE + RDC\cite{2K39}
        \item[d] Backrub method/Rosetta all-atom energy + RDC\cite{2K39}
        \item[e] MUMO method/CHARMM22 + NOE + RDC\cite{2NR2}
        \item[f] DER method/ CHARMM22 + NOE + $S^2$\cite{1XQQ}
        \item[g] NOE + RDC\cite{1D3Z}
        \item[h] X-ray 1.80 \AA\ structure\cite{1UBQ}
        \item[i] X-ray 1.80 \AA\ structure\cite{1UBI}
        \item[j] X-ray 1.32 \AA\ structure (synthetic protein with fluoro-LEU at residues 50 and 67)\cite{1OGW}
        \item[k] The methods presented here
    \end{tablenotes}
\end{threeparttable}
\end{table}

\end{document}